\documentstyle[12pt]{article}

\begin{document}

\begin{center}
CAUSAL PROPAGATORS FOR ALGEBRAIC GAUGES
\vspace{1cm}

B.M.Pimentel${}^{*}$, A.T.Suzuki${}^{\star}$, and J.L.Tomazelli${}^{\#}$ \\
\vspace{.1cm}
Instituto de F\'{\i}sica Te\'{o}rica \\
Universidade Estadual Paulista\\
Rua Pamplona, 145\\
01405-900 S\~{a}o Paulo, SP - BRAZIL\\

\vspace{3cm}
\end{center}

\vspace{1cm}

\begin{abstract}
Applying the principle of analytic extension for generalized funtions
we derive causal propagators for algebraic non-covariant gauges. The so
generated manifestly causal gluon propagator in the light-cone gauge is used
to evaluate two one-loop Feynman integrals which appear in the computation
of the three-gluon vertex correction. The result is in agreement with that
obtained through the usual prescriptions.
\end{abstract}

\vfill

Classification: {\em PACS. Nos. 11.15.Bt; 12.38.Bx}
\vspace{.5cm}

-------------------------------------------------------------------

${}^{*}$ {\footnotesize With partial support of CNPq, Bras\'{\i}lia}

${}^{\star}$ {\footnotesize E-mail: suzuki@axp.ift.unesp.br}

${}^{\#}$ {\footnotesize Supported by Capes, Bras\'{\i}lia}

\newpage

Over a quarter of a century ago Bollini, Giambiagi, and
Dom\'{\i}nguez${}^{[1,2]}$ considered causal distributions in
the context of the Fourier transform of radial functions,
$f(Q^2)$, where $Q^2\equiv k_{0}^{2} - \vec{k}^2 =
k_{0}^{2}-k_{1}^{2}-k_{2}^{2}-k_{3}^{2}$. By introducing a
positive parameter $\alpha $ such that
\begin{equation}
(f_{\alpha},\tau) = (f(\alpha ^{2}k_{0}^2-\vec{k}^2),\tau
(k_0,k_1,k_2,k_3)) \equiv \frac {1}{\alpha} (f(Q^2),\tau (\frac
{k_0}{\alpha},k_1,k_2,k_3)),
\end{equation}
where $\tau $ is a test function, one defines that $f_{\alpha}$
is analytic in $\alpha $ if for any $\tau $, the functional
$(f_{\alpha},\tau )$ is also analytic in $\alpha $.

Now, when $f_{\alpha }$ is analytically continued to the whole of
the upper half plane of $\alpha $, then a causal distribution is
defined through the following extension $k_{0} \rightarrow
\alpha k_{0}$, i.e.,
\begin{equation}
f(k^2 + i\epsilon ) = \lim_{\alpha \rightarrow 1+i\epsilon }
f(\alpha ^2 k_{0}^{2}-\vec{k}^2),\hspace {.5cm} \epsilon
\rightarrow 0^{+}.
\end{equation}

{}From this reasoning of analytic continuation as a postulate, one
can derive the covariant Feynman propagator in momentum space as
follows
\begin{equation}
\frac {1}{k^2} \rightarrow \lim_{\alpha \rightarrow 1+i\epsilon}
\frac {1}{\alpha ^2 k_{0}^{2}-\vec{k}^2} = \frac
{1}{k^2+2i\epsilon k_{0}^{2}}, \hspace{.5cm} \epsilon
\rightarrow 0^{+}.
\end{equation}

Since $k_{0}^{2}>0$, one has the usual prescription for handling
covariant poles, namely,
\begin{equation}
\frac {1}{k^2} \rightarrow \lim_{\varepsilon \rightarrow 0^{+}}
\frac {1}{k^2+i\varepsilon }, \hspace{.5cm} \varepsilon \equiv
2\epsilon k_{0}^{2} \rightarrow 0^{+}.
\end{equation}

Non--covariant (or algebraic) gauge choices which are
characterized by an external, constant vector $n_{\mu}$, on the
other hand leads to the appearance of gauge--dependent
poles\footnote{very often, incorrectly called ``unphysical'' or
``spurious'' poles.} $(k \cdot n)^{-\alpha}\,\,\,, \alpha = 1,
2, ...$ in the Feynman amplitudes. These therefore contain in
their structure, for instance, factors of the form
\[ \frac {1}{(k^2 + i\varepsilon)[k \cdot n]}\,\,\,\,,\]
where
\[\frac {1}{[k \cdot n]} \]
indicates that one still needs a matematically consistent way
of dealing with it at the pole $k\cdot n = 0$. Several
prescriptions have been defined and used in the literature for
this purpose. However, mathematics only does not suffice for
such a task as it has been demonstrated in the particular case
of the light--cone choice${}^{[3]}$; we also need to watch out
that causality is not violated by the prescription {\em per se}
or even in the process of its implementation in a direct
calculation. This is the reason why one should not consider
gauge--dependent poles as ``unphysical'', since they too must be
constrained by causality, so that in our field theory we will
not allow that positive--energy quanta propagating into the
future become mixed up with negative--energy ones propagating
into the past and vice--versa.

With the insight acquired in the work of reference [3], recently
Pimentel and Suzuki${}^{[4]}$ have proposed a causal
prescription for the light--cone gauge starting from the premise
that the propagator as a whole must be causal. In this sequel,
we propose that within the framework of analytic continuation as
discussed above, we can arrive at the very causal prescription
for the light--cone gauge. One notes, however, that
non--covariant poles (such as the light--cone one) are not of a
radial type function and no proof is given that such are
tempered distributions either. Yet, on the assumption that
analytic extension as defined above for the covariant pole is
legitimate and applicable to non--covariant poles, we draw some
interesting results.

To begin with, consider the product $(k^2 k\cdot n)^{-1}$ with
$n^{\mu}\equiv (n^0,0,0,n^3)$\footnote{for convenience we have
chosen components $n^1=n^2=0$} being an external, arbitrary
vector which determines the choice of a gauge of the algebraic or
non-covariant type. The factor $(k^2 k\cdot n)^{-1}$ upon the hypothesis of
analytic continuation becomes
\begin{equation}
\frac {1}{k^2 k\cdot n} \rightarrow \frac {1}{(k^2+2i\epsilon
k_0^2)(k\cdot n + i\epsilon k^{0}n^{0})} .
\end{equation}

As long as the external vector $n$ is quite arbitrary, we can
choose it so that  $n^{0}>0$, and since $\epsilon $ is strictly
positive, equation (5) may be rewritten as\footnote{recall that we
continue analytically to the whole of the {\it upper} half plane
of $\alpha $}
\begin{eqnarray}
\frac {1}{k^2 k\cdot n} & \rightarrow & \frac {1}{(k^2+2i\epsilon
k_0^2)(k\cdot n+i\epsilon \mid k^{0}\mid
n^{0})},\hspace{.5cm}for\hspace{.1cm} k^{0}>0 \nonumber \\
\frac {1}{k^2 k\cdot n} & \rightarrow & \frac
{1}{(k^2+2i\epsilon k_0^2)(k\cdot n-i\epsilon \mid k^{0}\mid
n^{0})},\hspace{.5cm}for\hspace{.1cm} k^{0}<0
\end{eqnarray}
or, using the Heaviside $\Theta $-function,
\begin{equation}
\frac{1}{k^2k\cdot n} \rightarrow \frac{1}{k^2+i\varepsilon
}\left\{ \frac {\Theta(-k^0)}{k\cdot n-i\xi}+\frac {\Theta
(k^0)}{k\cdot n+i\xi}\right\},\hspace{.5cm} \begin{array}{c} \varepsilon
\equiv 2\epsilon k_0^2 \rightarrow 0^{+} \\
\xi \equiv \epsilon \mid k^0 \mid n^0
\rightarrow 0^{+} \end{array},
\end{equation}
which is exactly the causal prescription considered in reference
[4].

We can, of course, generalize for higher order poles of $(k\cdot
n)$ as well as for non-covariant gauge choices other than the
light-cone one. We shall only consider the double pole case and
briefly discuss the pure homogeneous axial gauge $(n^0=0)$ and
the pure homogeneous temporal gauge $(n^3=0)$ choices for
$n^{\mu }$.

First of all, let us consider the simple pole cases. From
equation (5) we note that the analytic continuation of $k\cdot
n$ entails a sign dependence of the imaginary part coming from
$k^0$ and $n^0$.

The pure temporal case\footnote{we stick to the case $n^0>0$} is
such that there will be a violation of causality if one employs
the principal-value (PV) prescription to treat the pole
$(k\cdot n)^{-1}$, in the same manner as it breaks causality in
the light-cone case${}^{[3]}$. Indeed, evaluation of the Wilson
loop to the fourth order carried out by Caracciolo {\it et
al\/}${}^{[5]}$ has shown that in the temporal gauge the PV
prescription used to treat the gauge-dependent poles leads to
results which do not agree\footnote{not surprisingly since
causality has been broken by the PV prescription} with the ones
obtained in the Feynman and Coulomb gauges.

On the other hand, one has a very different situation for the
pure axial gauge for which $n^{\mu}=(0,0,0,n^3)$. Here, nothing
whatsoever can be said {\it a priori} whether a given
prescription to treat the gauge dependent pole $(k\cdot n)^{-1}$
will or will not violate causality, since analytic continuation
for this peculiar case is ill-defined.

Secondly, let us consider the double pole $(k\cdot n)^{-2}$
cases. In the light-cone gauge, the appearance or no of the double
pole factor $(k\cdot n)^{-2}$ in the Feynman amplitudes depends
upon whether one uses the four-component uneliminated formalism or the
eliminated two-component formalism. In the latter case, it
arises, for instance, in the evaluation of the one-loop gluon
self-energy. A sample calculation of a typical integral of this
type has been presented in reference [3], where use of the causal prescription
(or any other prescription which preserves causality) is {\it mandatory}.

Finally, a word on the axial gauge when one chooses for the
external vector $n^{\mu}=(n^0,0,0,n^3)$ such that
$(n^0)^2<(n^3)^2$. In this case, one {\it has to} use the causal
prescription (or any other prescription preserving causality) in
order to circumvent the gauge dependent poles.

We now proceed by implementing this causal vector boson propagator in two
types of integrals which occur in the evaluation of the ``swordfish''
diagrams of the three--gluon vertex correction when we employ the
two--component formalism of the light--cone gauge${}^{[7]}$. We shall see
that the outcome is concordant with the result obtained through the use of
the other prescriptions${}^ {[6]}$. In order to do this we follow previous
notation and conventions as employed in reference [7]. The integrals are:
\begin{equation}
{\cal K}(p,q) = \int \frac {d^{2\omega}r}{r^2 (r-q)^2}
\frac{1}{(p^{+}+r^{+})}
\end{equation}
and
\begin{equation}
{\cal K}^{l}(p,q) = \int \frac {d^{2\omega}r}{r^2 (r-q)^2} \frac
{(p^{l}+r^{l})}{(p^{+}+r^{+})}\,\,\,\,\,\, , \,\,\, l = 1,\, 2,
\end{equation}
which can be rewritten in a more convenient way as
\begin{equation}
\tilde{{\cal K}}(p,q) = \int \frac {d^{2\omega}r}{(r-p)^2(r-p-q)^2}
\frac {1}{r^{+}}\,\,,
\end{equation}
and
\begin{equation}
\tilde{\cal K}^{l}(p,q) = \int \frac
{d^{2\omega}r}{(r-p)^2(r-p-q)^2} \frac {r^{l}}{r^{+}}\,\,.
\end{equation}

The singularities at $r^{+} = 0$ are treated according to what
the principle of causality obliges for the whole of the boson
propagator, namely,
\begin{equation}
\frac {1}{r^2 r^{+}} \rightarrow \frac
{1}{r^2+i\varepsilon}\left [
\frac {\theta(r^{0})}{r^{+}+i\epsilon} + \frac
{\theta(-r^{0})}{r^{+}-i\epsilon}\right ]\,\,,
\end{equation}
where the infinitesimals $\varepsilon$ and $\epsilon$ goes to
zero from above, i.e., they are small, positive real numbers,
and $\theta(\pm x)$ is the usual Heaviside unit step function.
This propagator naturally ensures that positive--energy quanta
propagating into the future do not become mixed up with
negative--energy ones, and vice--versa into the past.

For the actual computation we decompose the momentum integration
into its longitudinal and transverse parts, so that for the
longitudinal part we can regularize the integral via dimensional
regularization in an Euclidean space of $2\omega -2$ dimensions
and for the transverse part we use
\begin{equation}
\frac {\theta (-r^0)}{r^{+}-i\epsilon}+\frac {\theta
(r^0)}{r^{+}+i\epsilon} = PV\frac{1}{r^{+}} - i\pi
\delta(r^{+})\frac{(r^{+}+r^{-})}{|r^{+}+r^{-}|}\,\,,
\end{equation}
with
\begin{equation}
PV\frac{1}{r^{+}}=\frac{1}{2}\left
[\frac{1}{r^{+}+i\epsilon}+\frac{1}{r^{+}-i\epsilon}\right ]\,\,,
\end{equation}

After some algebra, we arrive at the following partial results
\begin{equation}
\tilde{\cal K}(p,q) = i\frac {(-\pi)^{\omega}\Gamma
(2-\omega)}{(p^{+}+q^{+})}\left
\{(q^2)^{\omega-2}\int_{0}^{1}\,dy\,{\cal F}(y)-(\hat
q^{2})^{\omega-2}\int_{0}^{1}\,dy\,{\cal G}(y)\right \}\,\,,
\end{equation}
and
\begin{eqnarray}
\tilde{\cal K}^{l}(p,q) & = & (p^{l}+q^{l})\tilde{{\cal K}}(p,q)
   - i(-\pi)^{\omega}\Gamma(2-\omega)\frac {q^{l}}{p^{+}+q^{+}} \nonumber \\
&  &   \times \left\{(q^2)^{\omega-2}\int_{0}^{1}\,dy\,y\,{\cal
F}(y) - (\hat{q}^2)^{\omega-2}\int_{0}^{1}\,dy\,y\,{\cal G}(y)\right\}\,\,,
\end{eqnarray}
where
\begin{eqnarray}
{\cal F}(y) & \equiv & \frac {[y(1-y)]^{\omega-2}}{(1-\sigma y)}\,\,, \\
{\cal G}(y) & \equiv & \frac {[(y-\xi)(y-\bar{\xi})]^{\omega-2}}{(1-\sigma
y)}\,\,, \\
  \sigma & \equiv & \frac {q^{+}}{(p^{+}+q^{+})}\,\,, \\
  \xi  & \equiv & \frac {(1+\nu-\rho)+\sqrt{(1+\nu-\rho)^2-4\nu}}{2}\,\,, \\
  \bar\xi & \equiv & \frac {(1+\nu-\rho)-\sqrt{(1+\nu-\rho)^2-4\nu}}{2}\,\,, \\
  \nu & \equiv & \frac {2(p^{+}+q^{+})(p^{-}+q^{-})}{\hat q^2}\,\,, \\
  \rho & \equiv & \frac {2p^{+}p^{-}}{\hat q^2}\,\,, \\
  \hat q^2 & \equiv & q^{1}q^{1} +q^{2}q^{2} = 2q^{+}q^{-}-q^2\,\,, \\
  p^{\pm} & \equiv & \frac {(p^{0}\pm p^{3})}{\sqrt {2}}\,\,.
\end{eqnarray}

These results, Eqs. (15) and (16), agree with those of reference [7].
Moreover, in what follows we are going to explore a little
different pathway to evaluate de $y$--integrals. We first take
the limit $\omega \rightarrow 2$ and then perform the
$y$--integrations. Doing this enables us to obtain a ``better
looking'' form for the final results.

The end results for the integrals in question are:
\begin{equation}
\tilde{\cal K}(p,q)  =  \frac {i\pi^2}{q^{+}} T(p,q) +
{\cal O}(2-\omega)\,\,,
\end{equation}
where
\begin{eqnarray}
T(p,q) & \equiv & \ln (1-\sigma)\ln \left (\frac {q^2}{\nu \hat
q^2}\right ) + \ln \left (\frac {\xi}{\xi - 1}\right )\ln \left
(\frac {\sigma - 1}{\sigma \xi - 1}\right ) \nonumber \\
& & + \ln \left (\frac {\bar {\xi}}{\bar {\xi} - 1}\right )\ln
\left (\frac {\sigma - 1}{\sigma \bar {\xi} - 1}\right )
+ {\cal S}(\sigma) \nonumber \\
& & - {\cal S}\left (\frac {\sigma}{\sigma - 1}\right ) +
{\cal S}\left (\frac {\sigma \xi}{\sigma \xi - 1}\right ) +
{\cal S}\left (\frac {\sigma \bar {\xi}}{\sigma \bar {\xi} -
1}\right ) \nonumber \\
& & - {\cal S}\left (\frac {\sigma (\xi - 1)}{\sigma \xi -
1}\right ) - {\cal S}\left (\frac {\sigma (\bar {\xi} -
1)}{\sigma \bar {\xi} - 1}\right )\,\,,
\end{eqnarray}
and
\begin{equation}
\tilde {\cal K}^{l}(p,q)  =  \frac {(p^{l}q^{+} -
p^{+}q^{l})}{q^{+}}\tilde {\cal K}(p,q) - i\pi^{2} \frac
{q^{l}}{q^{+}}\,U(p,q) + {\cal O}(2-\omega)\,\,,
\end{equation}
where
\begin{equation}
U(p,q)  \equiv  \ln \left (\frac {q^2}{\nu \hat {q}^2}\right )
- (\xi - 1)\ln \left (\frac {\xi}{\xi -
1}\right ) - (\bar {\xi} - 1)\ln \left (\frac {\bar {\xi}}{\bar {\xi} -
1}\right )\,\,.
\end{equation}

In conclusion, we would like to emphasize and observe that we were able to
express Eqs. (26) and (28) in terms of products of logarithms
and in terms of various dilogarithms or Spence integrals, ${\cal
S}(\lambda)$, $\lambda$ being a general argument for the
dilogarithm. Although much more complex than the basic one--loop
light--cone integral, namely,
\begin{eqnarray}
{\cal K}(p) & \equiv &  \int \frac {d^{2\omega}\,r}{r^2 (r-p)^2
r^{+}}\nonumber \\
& = & \frac {i\pi^2}{p^{+}}\left \{\frac {\pi^2}{6} - {\cal
S}(\lambda)\right \} + {\cal O}(2-\omega)\,\,,
\end{eqnarray}
where, in this last equation, $\lambda $ stands for $\frac
{-\hat p^2}{p^2}$, Eqs. (26), (28), and (30) show us that basically
they belong to the same class of one--loop finite light--cone
integrals. The naive power counting to assess the degree of divergence of
these integrals remains valid.

\newpage

{\bf REFERENCES}
\vspace{0.7cm}

\begin{description}
\item[{[1]}] Bollini, C.G., Giambiagi, J.J., and
Dom\'{\i}nguez, A.Gonz\'alez {\it Journal of Math. Phys.} {\bf 6}, 165 (1965)

\item[{[2]}] Bollini, C.G., and Giambiagi, J.J. {\it Il Nuovo Cimento} {\bf
34}, 1146 (1965)

\item[{[3]}] Pimentel, B.M., and Suzuki, A.T. {\it Phys. Rev.} {\bf
D42}, 2115 (1990)

\item[{[4]}] Pimentel, B.M., and Suzuki, A.T. {\it Mod. Phys.Lett. A} {\bf
6}, 2649 (1991)

\item[{[5]}] S.Caracciolo, G.Curci, and P.Menotti, {\it Phys.Lett.} {\bf
113B}, 311 (1982)

\item[{[6]}] S.Mandelstam, S., {\em Nucl. Phys.} {\bf B 123}, 149 (1983);
G.Leibbrandt, {\em Phys. Rev.} {\bf D 29}, 1699 (1984)

\item[{[7]}] Suzuki, A. T., {\em  J. Math. Phys.} {\bf 29} 1032 (1988)
\end{description}
\end{document}